# light-scanning hand-held photoacoustic probe design

Yongjian Zhao[1,2,3], Luyao Zhu[1,2,3], Hengrong Lan[1,2,3], Daohuai Jiang[1,2,3], Feng Gao[1] and Fei Gao[1,*]

[1]Hybrid Imaging System Lab (HISLab), Shanghai Engineering Research Center of Intelligent Vision and Imaging, School of Information Science and Technology, ShanghaiTech University, Shanghai 201210, China

[2]Chinese Academy of Sciences, Shanghai Institute of Microsystem and Information Technology, Shanghai, China

[3]University of Chinese Academy of Sciences, Beijing, China

**Abstract**

**Significance:** We proposed a new design of hand-held linear-array photoacoustic (PA) probe which can acquire multi images via motor moving. Moreover, images from different locations are utilized via imaging fusion for SNR enhancement.

**Aim:** We devised an adjustable hand-held for the purpose of realizing different images at diverse location for further image fusion. For realizing the light spot which is more matched with the Ultrasonic transducer detection area, we specially design a light adjust unit. Moreover, due to no displacement among the images, there is no need to execute image register process. The program execution time be reduced, greatly.

**Approach:** mechanical design; Montel carol simulation; no-registration image fusion; Spot compression.

**Results:** Multiple PA images with different optical illumination areas were acquired. After image fusion, we obtained fused PA images with higher signal-to-noise-ratio (SNR) and image fidelity than each single PA image. A quantitative comparison shows that the SNR of fused image is improved by 36.06% in agar-milk phantom, and 44.69% in chicken breast phantom, respectively.

**Conclusions:** In this paper, the light scanning adjustable hand-held PA imaging probe is proposed, which can realize PA imaging with different illumination positions via adjusting the optical unit.

*Fei Gao, E-mail: gaofei@shanghaitech.edu.cn

## 1 Introduction

Nowadays, medical imaging technologies are developing very quickly, among which photoacoustic Imaging (PAI), an emerging non-invasive and non-ionizing imaging technique, attracts increasing attention in recent years [1]. In PAI, we detect the laser-induced pressure wave and perform image reconstruction [2]. It possesses both molecular contrast and high spatial resolution in deep tissue [3-5]. As for other traditional medical imaging methods: x-ray imaging involves ionizing radiation and non-optimal sensitivity as well as specificity [6, 7], magnetic resonance imaging (MRI) enhances contrast and spatial resolution with huge time consumption and high cost [8, 9]; ultrasonic imaging is non- invasive and low-cost, but the image quality requires enhancement and the false positive rate is high [1, 10]. Interestingly, PAT has the potential to overcome abovementioned drawbacks [11].

Due to its unique advantages, PAT system has been widely applied in hand-held detection scenarios, operative guided imaging (Biopsy pathological examination) [12], etc. These applications put forward some new requirements: 1) Moveable. Being portable ensures that doctors could adjust the detecting field of view (FOV) freely during the detection or operation. 2) Adjustable. It is expected to be possible to adjust the optical exciting unit or the ultrasonic receiving unit automatically by a control unit [13]. To address above needs, in this letter, we proposed an continuously-adjustable light-scanning PA probe for linear-array PA imaging system.

## 2  Hand-held Probe Design.

The overall design of the PA probe is shown in Figure 2(a)-(b). The PA probe can be divided into three parts: the first part is medical US linear-probe clamp (LPC) shown Figure 2(c). The second part is light transition unit (LTU) shown in Figure 2(d), which is made up of a cage flange, a right-angle prism and an optical diffuser & acoustic delay module made of polydimethylsiloxane (PDMS), aiming to transmit both light and ultrasonic waves. The third part is the optic wedge unit (OWU) for light beam shaping in Figure 2(e) (Material: SR11, refractive index: 1.843 at the wavelength of 532nm, wedge angle: $29°26'$) along with the mounting base, aiming at compressing the incident collimated light in one dimension to achieve the elliptical beam with the horizontal-to-vertical proportion of 3.41:1. Meanwhile, there are specific orbits on the wedge holder for the connection with the cage flange. We may obtain beam shaping units with different compressing ratios by changing OWU. The size of the PA probe is $60 \times 45 \times 80 mm^3$ (Shown in Figure 2(f)), and it can be held by an adult's single hand, as shown in Figure 1(a-b). For control the probe, we devised a control board based on Arduino platform consisting of power module, driver module and Voltage conversion module shown in Figure 1(c).

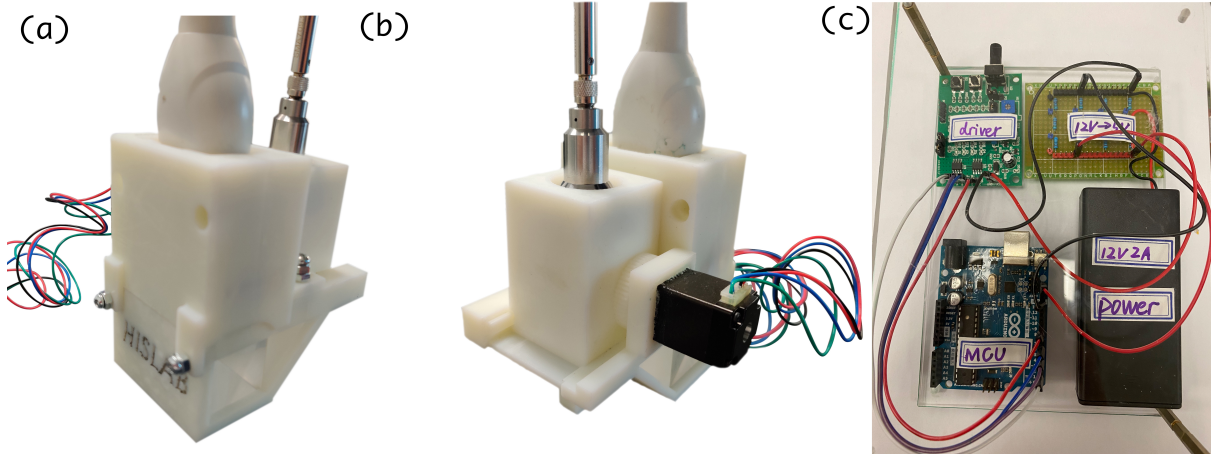

**Fig. 1** Photograph of the assembled handheld PA probe.(a-b) External structure of probe.(c)Control board.

An optically transparent Poly (vinylidene chloride) membrane covers the diffusing pad to protect the gel from desiccation and mechanical damage, facilitating relatively rough handheld operation. Accommodating to the handheld requirement, the free space light from the high-power pulsed laser (pulse duration: 7 ns, rep. rate: 8 Hz, wavelength: 532 nm, LPS-532-L, CNI) is coupled into a multi-mode fiber (FIB-1000-Duv-OSMA, Ideaoptics). After travailing through the collimator (F810SMA-543, Thorlabs, USA, NA: 0.39 in air), the divergent light turns into a collimated Gaussian-beam with the diameter of 20 mm and the divergence angle of 0.5 mm/rad. Afterwards, the beam travels through the shaping unit, with the major axis remaining to be 20 mm and minor axis compressed to be 5.45 mm. After being refracted twice, the beam reaches the sample surface and provides optical excitation. It is worth noting that OWU moving along the orbit is driven by a stepper motor fixed on the cage flange to provide rotational power input, where we connected OWU to a rack with a common mode with the gear. Moving OWU along the orbit, we can adjust the positions of beam excitation continuously, generating different illumination pattern of optical excitation for multiple PA image reconstruction.

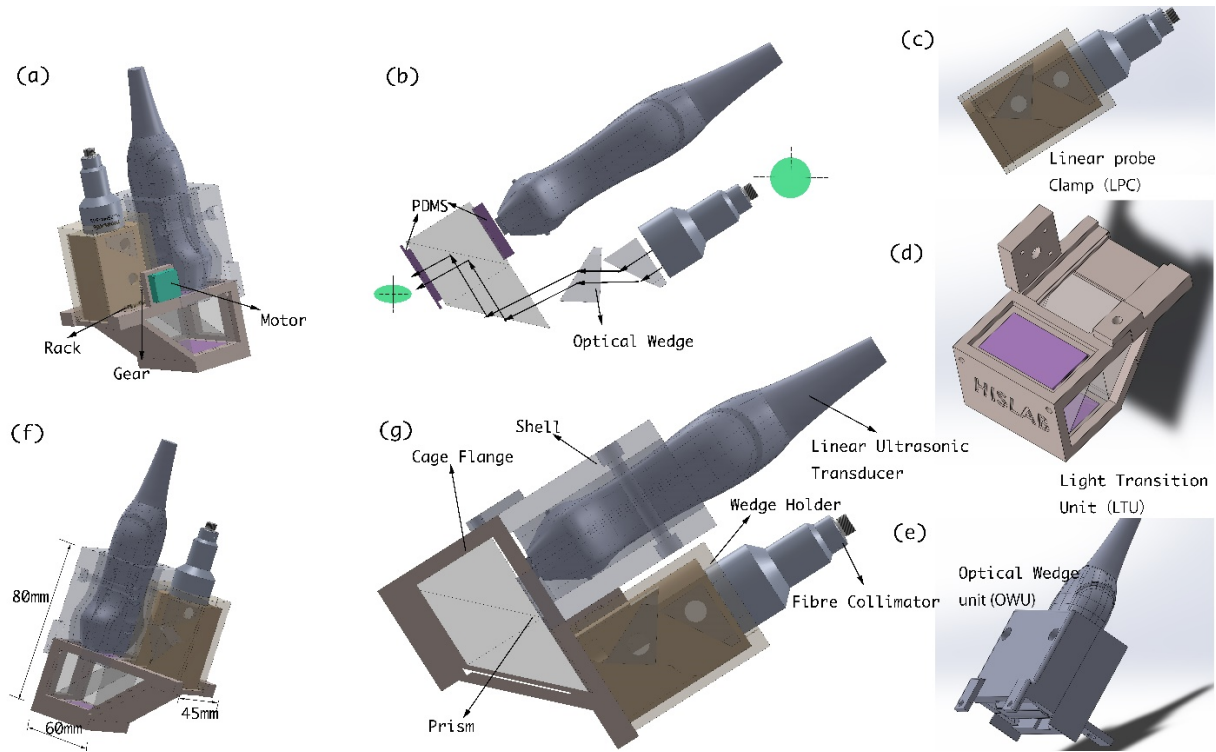

**Fig. 2** CAD (Computer Aided Design) photograph of the developed handheld probe for PA imaging. (a), (f) Axonometric of probe; (b) Schematic of beam shaping; (c)Linear Probe Clamp (LPC); (d)Light Transition Unit (LTU); (e) Optical Wedge Unit (OWU); (g) Front view of probe.

## 3　Numerical Simulation

To evaluate light illumination performance of the proposed probe, 3D fluence map within the region of interest (ROI) was estimated through the Monto Carlo simulation using Monte Carlo extreme (MCX) package[14]. We set the light source as an oval planar pattern, and there are two layers of simulating media: the upper one being PDMS (60×60×5 mm3) and the lower one being a tissue-mimicking medium (60×60×60mm3) for the propagation of light. The parameters of the PDMS layer are: refractive index n: 1.41@532nm, absorption coefficient: 0.001 mm-1, scattering coefficient: 0.1 mm-1. Parameters of the tissue-mimicking medium are: refractive index n: 1.35, absorption coefficient: 0.003 mm-1, scattering coefficient: 0.15 mm-1. Plus, we set two optical absorbers in the medium with the intervals of 20 mm in z-direction (cylinder, radius being 3mm, z-coordinates being 55 mm, 35 mm). The parameters of the two absorbers are: refractive index n: 1.54, absorption coefficient: 0.01 mm-1, scattering coefficient: 10 mm-1. By gradually scanning the position of the light source to simulate the working process of the probe, we obtained a set of optical fluence distribution images as shown in Figure 3.

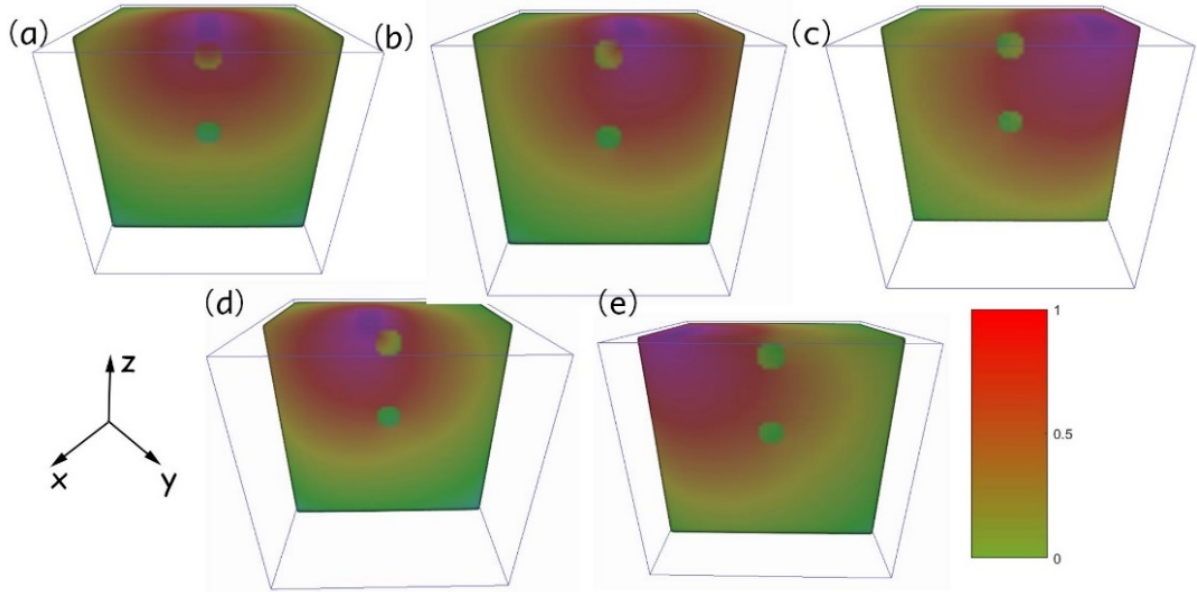

**Fig. 3** Simulation results of laser fluence at different illumination locations in tissue mimicking medium. (a) Light from directly above; (b)~(c) Light from the right; (d)~(e) Light from the left.

## 4　Experimental Setup

The experimental setup is shown in Figure 4. Q-switch pulsed laser is used for optical excitation with fiber coupling. After laser illumination on the sample, the induced PA wave passed through PDMS and prism layers, then reached the medical US linear array probe, followed by data acquisition (DAQ) using a US imaging system (S-Sharp) for future processing and image reconstruction. A personal computer synchronized the laser irradiation and acoustic data acquisition. The PA imaging speed was about 8 frames per second limited by the laser's repetition rate. The stepper motor is set to 16 subdivisions, which means that each pulse rotates 0.1125 degrees, and the motor speed is set to 60 r/min. There is a total of 6 detection positions, and the motor movement time between each position is 0.17 s, and the stop interval of each position is 3 s for image acquiring. We achieved 48 images for 6 positions spending 3.127 s on data acquiring, image reconstruction and image fusion. Therefore, it took a total of 18.977 s to perform the scanning process including motor movement and image processing. We can regard it as a real-time imaging probe with providing an imaging speed of 8 frames per second (limited by the repetition frequency of the laser). During the implementation of real-time imaging, we will consider the adjustable bright and dark fields of this probe more. Function, that is to say, researchers can choose bright field or dark field irradiation to meet experimental needs. For those experiments just for static imaging, our probe can realize the function of multi-picture fusion. This can provide more information on the same picture in the same experiment. The Figure 4 also demonstrates the relative position between the US detection area and the illumination area. The US area stay still while the illumination area moves with the OWU.

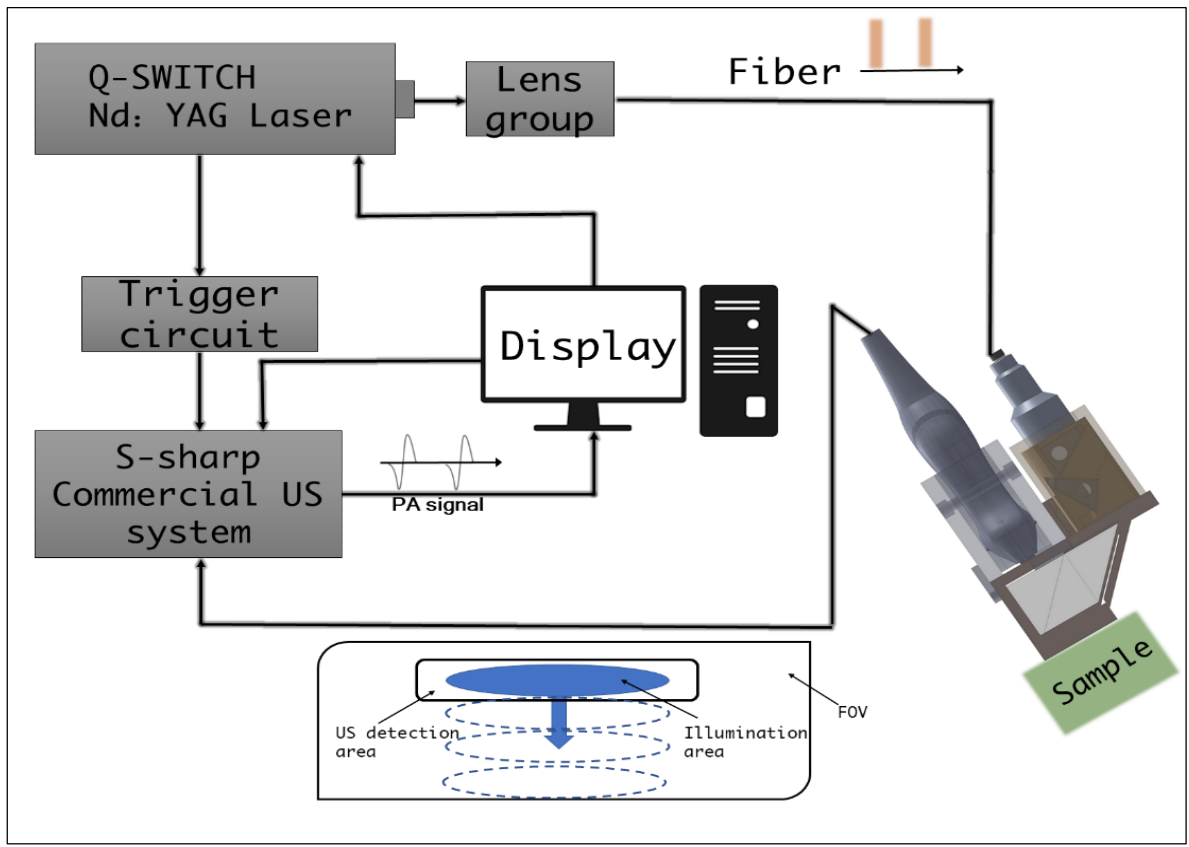

**Fig. 4** Schematic of the PA imaging system based on the proposed hand-held probe.

## 5 Experimental Study

### 5.1 Phantom study

We designed an agar-milk phantom for probe testing shown in Figure 5(a). Two cylinders were 3D printed with the diameters of 1 mm and lengths of 15 mm as optical absorbers (UV Curable

Resin), which were immerged into 5% black ink (absorption coefficient ~80 cm-1) for dying uniformly. Mixing 5g pure milk with 60 g distilled water to get the scattering medium, we boiled the solution with an induction cooker and added 2 g agar power at the same time, followed by stirring constantly for the acquisition of agar-milk solution. The two absorbers were vertically inserted in the solution, parallel to each other with the interval of 10 mm. The phantom was poured into the specific mold to solidify for the experiment. We perform two detection methods. One is that the probe detection area and the absorber are detected in parallel, and the other is that the probe and the absorber are placed vertically in space, to observe the detection result. The results are shown in Figure 6 and 7.

*5.2 Ex vivo study*

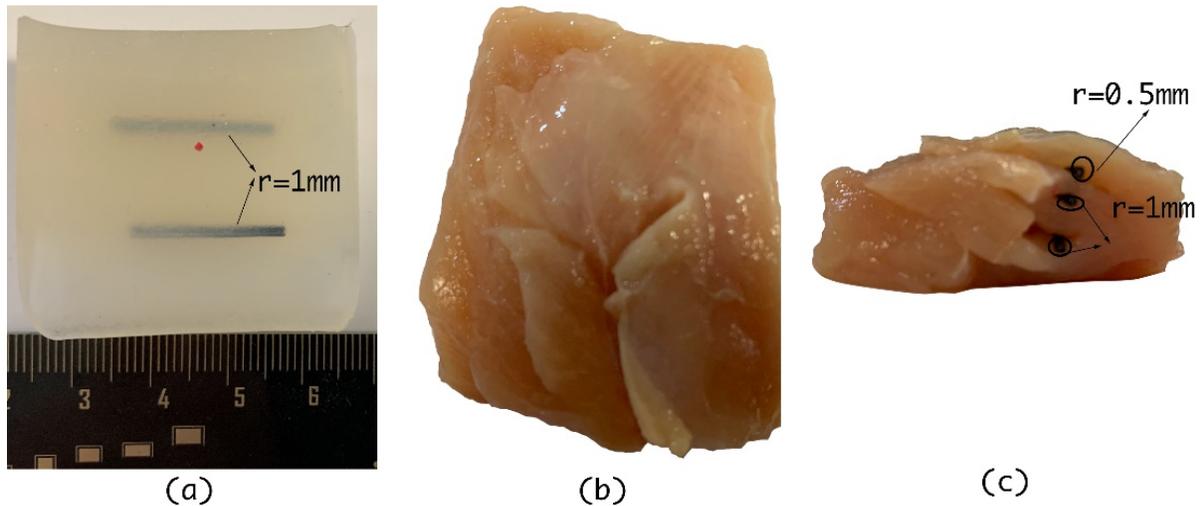

**Fig. 5** Photographs of phantom. (a) ager-milk phantom (b)~(c)chicken-breast phantom.

In order to further evaluate the performance of the probe, we used chicken breast for another phantom study. We picked up fresh chicken breast of 50g as the sample. The muscles at different depths were divided along the fiber texture with surgical forceps. Two different kinds of black rubber sticks were selected as the optical absorbers, diameters being 0.5 mm and 1 mm, respectively. The 0.5 mm stick was set on the uppermost layer, while the two 1 mm sticks were set on the two lower layers, each being 5 mm apart from the upper one. Afterwards, the muscles were cautiously rehabilitated, and the fabricated sample was shown in Figure 5(b)~(c).

## 6  Results

We conducted experiments with both agar phantom and chicken breast phantom. The probe was held by hand, and the samples were attached to the detection region coated with US coupling gel. Afterwards, we started the motor for controlled scanning. The acquired data were loaded into MATLAB for image reconstruction and fusion, with no need for image registration since the relative position of probe and samples was unchanged. The imaging result of agar phantom was shown in Figure 6, while those of chicken breast phantom and fused image were shown in Figure 7 and Figure 8. The last two pictures of experiment were abandoned. The remained four pictures were used to characteristic the results of PA imaging.

### 6.1  Phantom results

As shown in Figure 6 and Figure 7, when the light was directly above the samples (Figure 6&7(a)), the upper absorber was clearly demonstrated, while the lower one showed very weak PA signals since it received much fewer photons due to the shielding of upper absorber. As the light spot moved away from the confocal zone, PA signals of the upper absorber reduced gradually, while that of the lower one is enhanced (Figure 6&7 (b)-(c)). The reason is that as the relative displacement between the beam and absorber increased, the upper absorber receives less and less photons. On the other hand, the lower one might receive more light. Equivalently, the illumination pattern is gradually switching from bright-field to dark-field illumination. When the illumination arrives the boundary of sample, as shown in Figure 6&7 (d), the lower absorber generates stronger PA signal due to more photons irradiation.

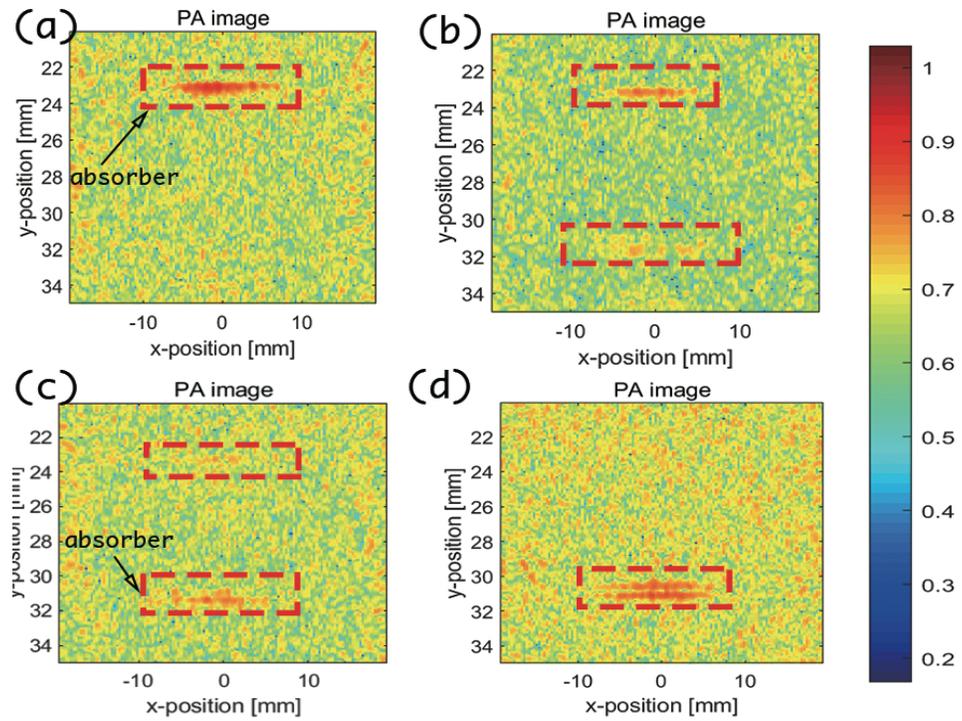

**Fig. 6** Ager-milk phantom imaging results with different illumination locations in parallel detection mode: (a) light on top the center of sample, (b) –(d)1cm, 2 cm, and 3 cm away from the center zone.

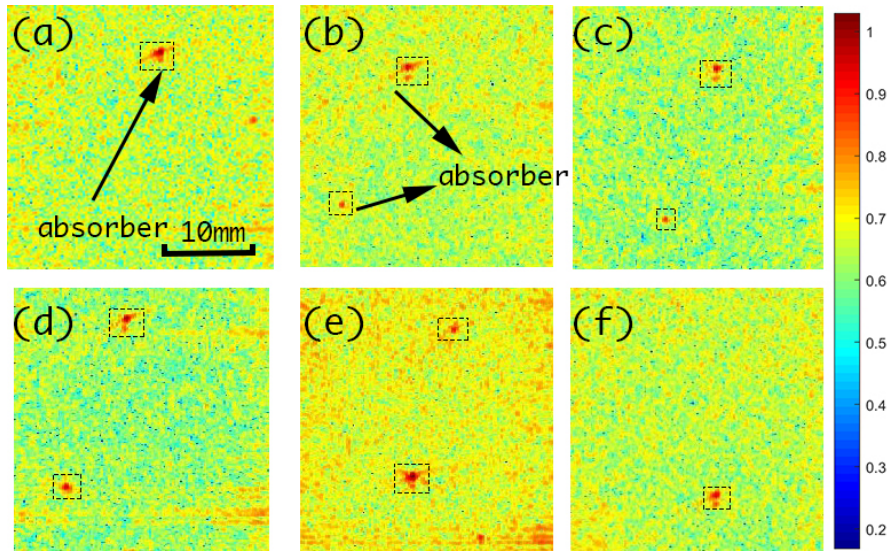

**Fig. 7** Ager-milk phantom imaging results with different illumination locations in vertical detection mode: (a) light on top the center of sample, (b) - (d)1cm, 2 cm, and 3 cm away from the center zone. (e) - (f) 3.5cm and 4 cm away from the center zone.

*6.2 Ex vivo experiment results*

Figure 8 shows the chicken breast imaging results, where we obtained 4 images showing the positions and outlines of the three absorbers. The lowest absorber shows weakest signals due to its large depth, and the middle absorber demonstrated the strongest signals in Figure 7(a) and (b). As the light spot moved, the middle absorber's signal decreased while the lowest one is slight enhanced. We may conclude that we can acquire PA images with different emphases through the light scanning mechanism, which provides more details.

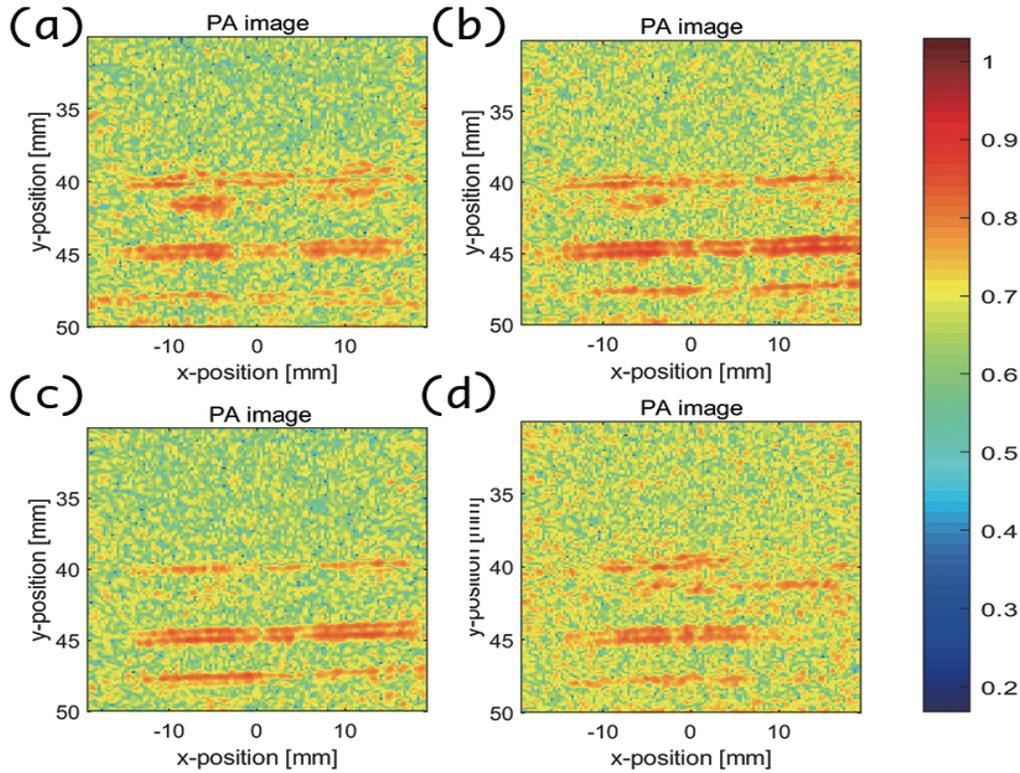

**Fig. 8** Chicken-breast phantom results at different time points corresponding to different illumination locations: (a) light on top the center of sample, (b) –(d)1cm, 2 cm, and 3 cm away from the center zone.

## 6. *Image fusion method and results*

According to the different ways of applying image information processing, image fusion can be divided into three levels of research, namely pixel level, feature level and decision level. Among them, the pixel-level fusion is located at the lowest layer, which can be regarded as only extracting the information and using it directly. It is also because of its maximum retention of information that it is superior to the other two levels in accuracy and robustness. In contrast, pixel-level fusion obtains more detailed information and is the most commonly used fusion method. At the same time, because there is no sensor displacement in our image acquisition process, there is no registration problem, so it can greatly reduce the time-consuming problems caused by registration. In this paper, we use the average method for image fusion. And calculate the signal-to-noise ratio of the image. The formula of SNR can be stated as :

$$SNR(db) = 10 log_{10} \left[ \frac{\sum_{x=1}^{N_x} \sum_{y=1}^{N_y} (f(x,y))^2}{\sum_{x=1}^{N_x} \sum_{y=1}^{N_y} \left(f(x,y) - \overline{f}(x,y)\right)^2} \right] \quad (1)$$

After image fusion, Figure 8(a) shows agar-milk imaging results with SNR's growth up to 50.94%. Similarly, the image quality of chicken breast phantom is also clearly improvement, and the SNR's enhancement was up to 30.93% compared with single PA image. More importantly, all the absorber can be clearly identified in the fused image with high contrast. Moreover, since there was no relative displacement for the US transducer and sample, no image registration was need, guaranteeing easy operation and computation cost.

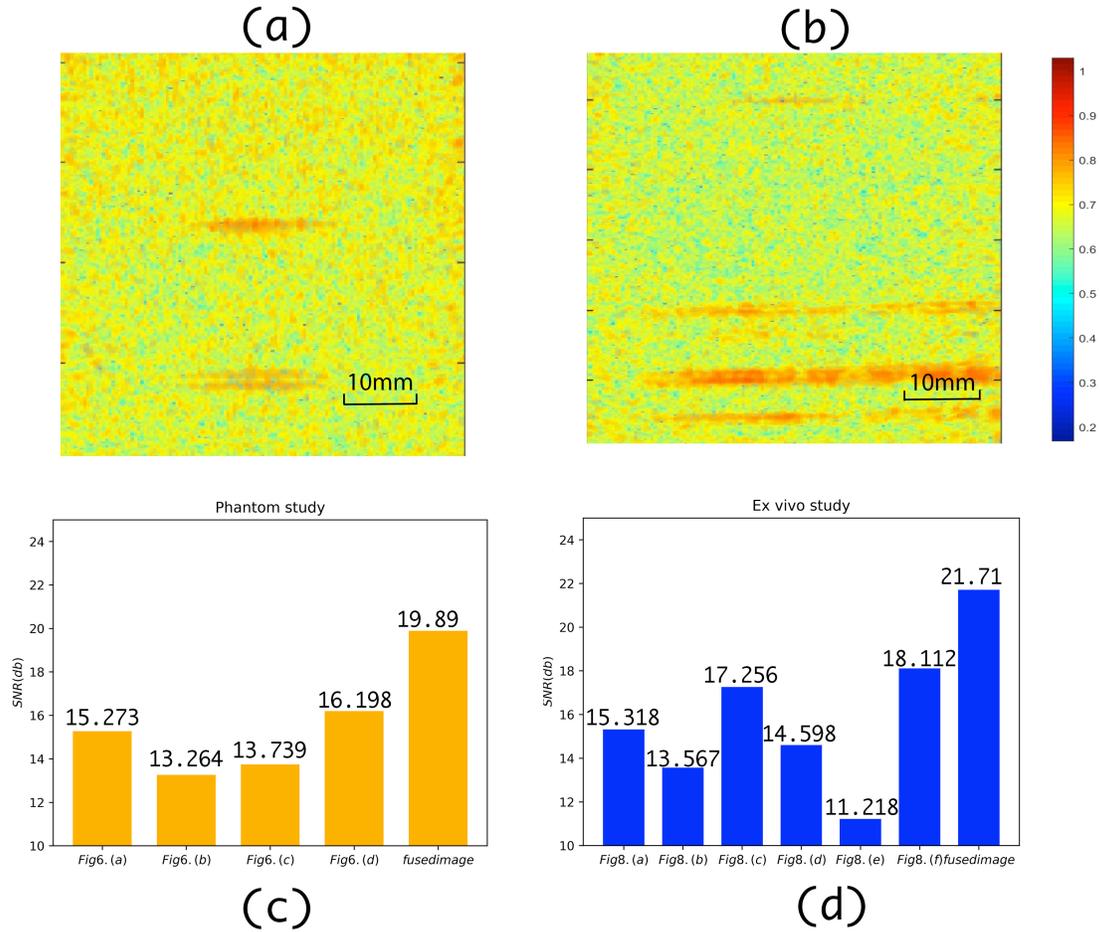

**Fig. 9** Fused images for both experiments: (a) agar-milk phantom, (b) chicken breast phantom, and (c)quantitative evaluation of PA images before and after image fusion.

## 7  Conclusion and discussion

In this paper, the light scanning adjustable hand-held PA imaging probe is proposed, which can realize PA imaging with different illumination positions via adjusting the optical unit. To achieve matched beam shape with US detection area, a couple of optical wedges were used to realize the compressing ratio of 3.41:1. The MCX simulation results revealed that the absorbers of different depth would receive different amounts of photons. And this proposed probe can adjust the illumination area with no change of relative position between the US transducer and sample. Both agar-milk and chicken breast phantom imaging results showed the position and contour of the absorbers well. Moreover, by image fusion, the SNR of the fused image is increased by 36.06% and 44.69%, respectively compared to the average SNR of the pre-processed image. Experimental results demonstrate the high feasibility and potential for clinical applications.

## 8 Disclosure

Fei Gao and Yongjian Zhao report patents, which are owned by the ShanghaiTech University and licensed to commercial entities, that are related to the technology and analysis methods described in this study. The authors have no relevant financial interests in this article and no potential conflicts of interest to disclose.

**Yongjian Zhao** received his bachelor's degree in Mechanical Engineering at Anhui University of Technology in 2018, Anhui, China. He is currently working toward the Master's degree at ShanghaiTech University, Shanghai, China. He has been awarded the top national scholarship in 2015 during his undergraduate period. His research interests include the biomedical and clinical device design, and super-resolution photoacoustic imaging based on hand-held probe, trying his best to combine the mechanical design and biomedical application.

**Caption List**

**Fig. 1** Photograph of the assembled handheld PA probe. (a-b) External structure of probe.(c)Control board.

**Fig. 2** CAD (Computer Aided Design) photograph of the developed handheld probe for PA imaging. (a), (f) Axonometric of probe; (b) Schematic of beam shaping; (c)Linear Probe Clamp (LPC); (d)Light Transition Unit (LTU); (e) Optical Wedge Unit (OWU); (g) Front view of probe.

**Fig. 3** Simulation results of laser fluence at different illumination locations in tissue mimicking medium. (a) Light from directly above; (b)~(c) Light from the right; (d)~(e) Light from the left.

**Fig. 4** Schematic of the PA imaging system based on the proposed hand-held probe.

**Fig. 5** Photographs of phantom. (a) ager-milk phantom (b)~(c)chicken-breast phantom.

**Fig. 6** Ager-milk phantom imaging results with different illumination locations in parallel detection mode: (a) light on top the center of sample, (b) –(d)1cm, 2 cm, and 3 cm away from the center zone.

**Fig. 7** Ager-milk phantom imaging results with different illumination locations in vertical detection mode: (a) light on top the center of sample, (b) - (d)1cm, 2 cm, and 3 cm away from the center zone. (e) - (f) 3.5cm and 4 cm away from the center zone.

**Fig. 8** Chicken-breast phantom results at different time points corresponding to different illumination locations: (a) light on top the center of sample, (b) –(d)1cm, 2 cm, and 3 cm away from the center zone.

**Fig. 9** Fused images for both experiments: (a) agar-milk phantom, (b) chicken breast phantom, and (c)quantitative evaluation of PA images before and after image fusion.